\documentclass[aps,rmp,amsmath,amsfonts,amssymb,superscriptaddress,floatfix,nofootinbib,notitlepage]{revtex4-1}
\usepackage{times}
\usepackage{graphicx}
\usepackage{enumerate}
\usepackage{natbib}
\usepackage{bm}
\usepackage{xcolor}
\definecolor{darkbrown}{rgb}{0.57,0.28,0.15}
\usepackage[breaklinks=true]{hyperref}
\hypersetup{
allcolors       = {blue},
linkbordercolor = {white},
linkcolor       = {cyan},
urlcolor        = {darkbrown},
colorlinks      = true
}

\usepackage{xcolor}

\newcommand{\av}[1]{\left\langle #1\right\rangle}  

\newcommand{\sign}{\mathop{\text{sign}}} 
\renewcommand{\Im}{\mathop{\mathrm{Im}}} 
\renewcommand{\Re}{\mathop{\mathrm{Re}}} 
\newcommand{\bnabla}{\bm{\nabla}}        

\newcommand{\defi}{{:=}}                            
\newcommand{\dd}{\mathrm{d}}                        
\newcommand{\ii}{\mathrm{i}}                        
\newcommand{\ee}{\mathrm{e}}                        
\newcommand{\bsigma}{\bm{\sigma}}        
\newcommand{\ba}{\mathbf{a}}             
\newcommand{\bb}{\mathbf{b}}             
\newcommand{\bk}{\mathbf{k}}             
\newcommand{\bl}{\mathbf{l}}             
\newcommand{\bbm}{\mathbf{m}}            
\newcommand{\bn}{\mathbf{n}}             
\newcommand{\bq}{\mathbf{q}}             
\newcommand{\br}{\mathbf{r}}             
\newcommand{\bu}{\mathbf{u}}             
\newcommand{\bv}{\mathbf{v}}             
\newcommand{\bA}{\mathbf{A}}             
\newcommand{\bL}{\mathbf{L}}             
\newcommand{\bP}{\mathbf{P}}             
\newcommand{\hb}{\widehat{b}}  
\newcommand{\hk}{\widehat{k}}  
\newcommand{\hq}{\widehat{q}}  
\newcommand{\bhb}{\mathbf{\widehat{b}}}  
\newcommand{\bhk}{\mathbf{\widehat{k}}}  
\newcommand{\bhq}{\mathbf{\widehat{q}}}  
\newcommand{\bhr}{\mathbf{\widehat{r}}}  
\newcommand{\bhs}{\mathbf{\widehat{s}}}  
\newcommand{\sfz}{\mathsf{z}}             
\newcommand{\sfG}{\mathsf{G}}             
\newcommand{\sfI}{\mathsf{I}}             
\newcommand{\sfN}{\mathsf{N}}             
\newcommand{\avphi}[1]{\av{#1}_{\!\!\phi}}

\begin{document}
\markboth{Y.-P. Pellegrini}{Uniformly-moving non-singular dislocations with elliptical core shape in anisotropic media}
\title{Uniformly-moving non-singular dislocations with elliptical core shape in anisotropic media}
\author{Yves-Patrick \surname{Pellegrini}}
\email{yves-patrick.pellegrini@cea.fr}
\affiliation{CEA, DAM, DIF, F--91297 Arpajon, France.}

\begin{abstract}
To allow for ``relativistic''-like core effects, an anisotropic regularization of steadily-moving straight dislocations of arbitrary orientation is introduced, with two scale parameters $a_\parallel$ and $a_\perp$ along the direction of motion and transverse to it, respectively. The dislocation core shape is an ellipse. When $a_\perp/a_\parallel\to 0$, the model reduces to the Peierls-Eshelby dislocation, the fields of which are non-differentiable on the slip plane. For finite $a_\parallel$ and $a_\perp$, fields are everywhere differentiable. Applying the author's so-called ``causal'' Stroh formalism to the model, explicit expressions for the regularized fields in anisotropic elasticity are derived for any velocity. For faster-than-wave velocities, Mach-cone angles are found insensitive to the ratio $a_\parallel/a_\perp$, as must be. However, the larger $a_\parallel$, the weaker the intensity of the cone branches. An expression is given for the radiative dissipative force opposed to motion. From this expression, it is inferred that the concept of a ``radiation-free'' intersonic velocity can, when not applicable, be replaced by that of a ``least-radiation'' velocity.
\end{abstract}
\maketitle

\section{Introduction}
This work is concerned with the regularization of dislocation fields, for a straight dislocation moving steadily in an anisotropic medium \citep{SAEN53,BULL54,TEUT62a}. Indeed, the so-called \emph{standard model} of dislocations \citep{ANDE17}, based on zero-width Volterra dislocations, produces singular fields at the origin. Various models have been proposed to tame this singularity. The oldest one is perhaps the Peierls-Nabarro model of a static dislocation \citep{PEIE40,NABA47,SCHO05b}, which was extended to steady motion \citep{ESHE49b,ESHE53b,WEER69a,WEER69b,WEER80a,MARK01c,ROSA01,PELL14}. However, as pointed out by Eshelby \citep{ESHE49b}, such Somigliana dislocations generate non-differentiable fields on the slip plane.

Since, a variety of regularization techniques were used to make the fields partially or fully nonsingular, mostly in a static context. Early models have been reviewed by \citet{CAIA06}, who put forward a phenomenological model of a dislocation with isotropic core of finite size, applying it to static dislocation segments. Another isotropic-core model has been proposed by \citet{PELL15} to compute the elastodynamic fields of arbitrarily-moving dislocations in an isotropic medium \citep{LAZA16a}. Also, non-locality is an efficient regularizing device. In this line of thought, non-local models have been introduced in the framework of gradient elasticity, such as a 6-parameter anisotropic core model, from which non-singular static fields in anisotropic crystals have been derived \citep{LAZA15a,LAZA15b,POLA18}.

The present work describes an extension of the Pellegrini-Lazar model to \emph{elliptical cores}, with application to steady motion in an anisotropic medium. The aim is to devise a simple model with two different regularizing length scales ---one in the transverse direction, and  the other one in the direction of motion--- so as to allow for `relativistic' core-width variations in the latter direction \citep{ESHE49b}. Such variations, which can be considered as the elastic counterpart of a Lorentz-Fitzgerald contraction, are important for motions of velocity comparable to wavespeeds in the crystal, including faster-than-wave velocities \citep{BULL54,ESHE56a,WEER67b,WEER69b,ROSA01,LAZA09a,PELL12,PELL14}; see, e.g., \citet{AULD73} for waves in crystals. However, the two length scales are considered as fixed parameters hereafter, this work being focused on obtaining explicit field expressions.

The paper is organized as follows. Section \ref{sec:eshdis} introduces the elliptical-core model, and shows that it can be considered a fully-regularized anisotropic extension of the Peierls-Eshelby model of a dislocation \citep{PEIE40,NABA47,ESHE49b}. Differences with the Lazar-Po approach \citep{LAZA15a,LAZA15b} are pointed out. The elastic fields of the moving dislocation are derived in Section \ref{sec:gsetup}. The method used relies on the Stroh formalism for dislocations moving through an anisotropic medium \citep{STRO62a,ANDE17,CHAD77,BACO79,LOTH92d,WU00,BARN02a,BLAS18}. This formalism was modified in a causal way in \citet{PELL17} to provide a general framework for field computations at any velocity, including faster-than-wave ones. From these results the dissipative force acting on the dislocation is obtained. Section \ref{sec:isolim} is devoted to the isotropic limit of the model. Numerical results are presented in Section \ref{sec:nures}. Section \ref{sec:concl} concludes the article. An Appendix gathers the most technical calculations.

Our conventions for the Fourier transform of a space- and time-dependent function $f(\br,t)$ are
\begin{align}
f(\bk,\omega)=\int\dd^d\!x\,\dd t\,f(\br,t)\ee^{-\ii(\bk\cdot\br-\omega t)},\,
f(\br,t)=\int\frac{\dd^d\!k}{(2\pi)^d}\frac{\dd \omega}{2\pi}\,f(\bk,\omega)\ee^{\ii(\bk\cdot\br-\omega t)},
\end{align}
where integrations with respect to the space variable $\br$ and the wave vector $\bk$ are over the $d$-dimensional space ($d=2$), and integrations with respect to the time $t$ and the angular frequency $\omega$ are over $\mathbb{R}$. Bold and sans-serif typefaces are used, respectively, to denote vectors ($\ba$) and dyadic tensors ($\mathsf{A}$) of components $A_{ij}$. A hat denotes a unit vector whenever it is built from an existing non-unit vector, e.g., $\bhk\equiv \bk/k$. This `hat' notation is not used for vectors specifically introduced as unit vectors. Finally, matrix or vector scalar products are denoted by a dot, which represents a contraction with respect to the tensor index in between when no confusion is possible: $\mathsf{a}\cdot\mathsf{b}$ is the matrix of components $a_{ik}b_{kj}$.

\section{Regularization: a nested family of dislocation models}
\label{sec:eshdis}
This section introduces a nested family of dislocation models rooted in the Peierls model. The dislocation is represented by the plastic eigenstrain tensor $\beta^{\rm p}_{ij}(\br,t)$ \citep{MURA87}. The dislocation, moving at constant velocity $\bv$, is such that in the direct and Fourier representations (respectively),
\begin{align}
\label{eq:betaexpr0}
\beta_{ij}^{\rm p}(\br,t)&=\beta_{ij}^{\rm p}(\br-\bv t),\qquad
\beta_{ij}^{\rm p}(\bk,\omega)=(2\pi)\delta(\omega-\bv\cdot\bk)\beta_{ij}^{\rm p}(\bk).
\end{align}
Its shape is assumed rigid (i.e., time-independent), and completely characterized by $\beta_{ij}^{\rm p}(\br)$, or by $\beta_{ij}^{\rm p}(\bk)$ in the Fourier representation. For steady-state motion, all quantities in direct space depend only on the position vector $\br-\bv t$ with respect to the dislocation position $\bv t$, where $\bv$ is the velocity. Hereafter, this vector is simply denoted by $\br$ for conciseness, all calculations being done in the co-moving frame. We introduce orthogonal unit vectors $\bbm$ and the slip plane normal $\bn$ along the $x$ and $y$ coordinate axes, respectively. Motion is along $\bbm$ so that $\bv=v\,\bbm$, where $v$ is the \emph{signed} scalar velocity. The dislocation line is orthogonal to both $\bbm$ and $\bn$.

With $\bb$ the Burgers vector, the plastic eigenstrain reads $\beta^{\rm p}_{ij}(\br)=n_i\hb_j\eta(\br)$, where $\eta(\br)=\eta(x,y)$ is the slip function along $x$. As it is localized on the slip plane, $\eta(\br)$ is a peaked function of $y$ near to $y=0$. It can be written as the convolution of a dislocation-density component $\rho(\br)=-\partial_x\eta(\br)$ by the unit-step Heaviside function $\theta$, as
\begin{align}
\eta(\br)&=\int_{-\infty}^{+\infty}\dd x'\theta\bigl(-(x-x')\bigr)\rho(x',y).
\end{align}
In the Fourier representation, the above expressions become
\begin{align}
\label{eq:etak}
\beta^{\rm p}_{ij}(\bk)&=n_i\hb_j\eta(\bk),\qquad \eta(\bk)=\frac{\ii\,\rho(\bk)}{\bk\cdot\bbm+\ii 0^+}.
\end{align}
Bearing in mind as a reference the singular Volterra model of a point-like dislocation characterized by
\begin{align}
\rho(\br)&=b\delta(x)\delta(y),\quad \rho(\bk)=b,\quad \eta(\br)=b\theta(-x)\delta(y),
\end{align}
the regularized models considered in this work are as follows:
\begin{enumerate}[(i)]
\item Model III: this new model is introduced to represent a dislocation with different half-widths $a_\parallel$ and $a_\perp$ in both directions.

With the diagonal matrix $\sfz=a_\parallel\bbm\otimes\bbm+a_\perp\bn\otimes\bn$, it reads
\begin{align}
\label{eq:pell}
\rho(\br)&=\frac{b}{2\pi\det\sfz}\frac{1}{\left[1+\br\cdot\sfz^{-2}\cdot\br\right]^{3/2}},\quad
\rho(\bk)=b \ee^{-\sqrt{\bk\cdot\sfz^2\cdot\bk}},\nonumber\\
&{}\hspace{2cm}\eta(\br)=\frac{b}{2\pi a_\perp}
\frac{1-(x/a_\parallel)/\sqrt{1+\br\cdot\sfz^{-2}\cdot\br}}{1+(y/a_\perp)^2}.
\end{align}

\item Model II: The Pellegrini-Lazar isotropic-core model \citep{PELL15}. It is a particular case of \eqref{eq:pell} for $a_\parallel=a_\perp=a$, and $\sfz=a\,\sfI$.

\item Model I: The semi-regularized Peierls-Eshelby dislocation model \citep{ESHE53b,ROSA01,MARK01b,MARK01c,PELL10b,PELL12,PELL14,PELL17}, characterized by
\begin{align}
\rho(\br)
&=\frac{b}{\pi a}\frac{\delta(y)}{1+(x/a)^2},\quad
\rho(\bk)=b\ee^{-a |k_x|},\quad
\eta(\br)=\frac{b}{\pi}\left(\frac{\pi}{2}-\arctan\frac{x}{a}\right)\delta(y).
\end{align}
where the length $a$ is the half core width. It is regularized along $x$ but its fields are non differentiable on the slip plane $y=0$.
\end{enumerate}

Models II and III are nonsingular. Relationships between models are obvious from $\rho(\bk)$. All three models reduce to the Volterra model when $a$, $a_\parallel$, and $a_\perp\to 0$ simultaneously. Also, the integral over the whole $y$ axis of $\eta(x,y)$ (i.e., the $k_y=0$ Fourier mode) gives the same result in Models I, II, and III upon taking $a_\parallel=a$. Finally, model III reduces to model I in the limit $a_\perp\to 0$ if $a_\parallel=a$. All models can thus be considered as members of the same family.

Hereafter, derivations are done for model III, which is no more complicated than model II, except for a few rescalings (in particular, the integrals to be considered are, technically, \emph{exactly} the same).

The density $\rho(\bk)$ in \eqref{eq:pell}${}_2$ is akin to some Fourier dual of that considered by Lazar and Po, who use (in the above notations) the exponential form $\rho(\br)\propto\exp{-\sqrt{\br\cdot\sfz^{-2}\cdot\br}}/$ $[4\pi\det(\sfz)\sqrt{\br\cdot\sfz^{-2}\cdot\br}]$ as an anisotropic density in the direct space, which is the Green function of the anisotropic Helmholz equation \citep{LAZA15a,LAZA15b}. By contrast, the real-space expression $\rho(\br)$ in \eqref{eq:pell}${}_1$ is finite at $r=0$. Note that, as can be seen from its Fourier transform, it is the solution of the following formal integro-differential equation, which involves the fractional anisotropic Laplacian $(-\bnabla\cdot\sfz^2\cdot\bnabla)^{1/2}$ (the one-dimensional version of which is the integral operator of the Peierls model  \citep{JOSI18}):
\begin{align}
\label{eq:diffeq}
\ee^{\sqrt{-\bnabla\cdot\sfz^2\cdot\bnabla}}\rho(\br)&=b\,\delta(\br).
\end{align}

\section{Elastic distortion, stresses, and forces in Model III}
\label{sec:gsetup}
\subsection{General setup}
The method of computation of the fields in Model III parallels that for Model I in \citet{PELL17}, to which the reader is referred for details. With $\bu$ the material displacement vector field, the elastic distortion $\beta_{ij}\defi\partial_i u_j-\beta^{\rm p}_{ij}$ is the main field of interest. It is computed by Fourier inversion as
\begin{align}
\label{eq:betaB}
\beta_{ij}(\br)
&=\int\frac{\dd^2\!k}{(2\pi)^2}\left[k_i G^+_{jp}(\bk,\bk\cdot\bv) k_q c_{pqkl}-\delta_{ik}\delta_{jl}\right]\beta^{\rm P}_{kl}(\bk)\ee^{\ii \bk\cdot\br}\nonumber\\
&=\ii\,b_l\int\frac{\dd^2\!k}{(2\pi)^2}\left[k_i G^+_{jp}(\bk,\bk\cdot\bv) k_q c_{pqkl}-\delta_{ik}\delta_{jl}\right]\frac{n_k \ee^{\ii \bk\cdot\br}\ee^{-\sqrt{\bk\cdot\sfz^2\cdot\bk}}}{\bk\cdot\bbm+\ii 0^+},
\end{align}
where $G^+_{jp}(\bk,\omega)$ denotes the Fourier expression of the elastodynamic retarded (i.e., causal) Green operator, where Eqs.\ \eqref{eq:betaexpr0}${}_2$, \eqref{eq:etak}, and \eqref{eq:pell}${}_2$, have been used, and where the immediate Fourier inversion with respect to $\omega$ has been carried out. The Green operator has the well-known expression
\begin{eqnarray}
\label{eq:fgdef}
\sfG^+(\bk,\omega)=\lim_{\epsilon\to 0^+}[\sfN(\bk)-\rho(\omega+\ii\epsilon)^2\sfI\,]^{-1},
\end{eqnarray}
where $N_{ij}(\bk)\defi k_k c_{iklj}k_l$ is the acoustic tensor built from the elastic tensor $c_{ijkl}$, and $\sfI$ is the $3\times3$ identity matrix. The retarded character of $\sfG^+$ stems from the shift of the real angular frequency $\omega$ to complex values with infinitesimal positive imaginary parts, which is a radiation condition \citep{BART89}. The non-trivial limit makes $\sfG^+(\bk,\omega)$ a generalized function (distribution).  The radiation condition in $\sfG^+(\bk,\bk\cdot\bv)$ can be transferred to $\bv$ by introducing the complex velocity vector
\begin{align}
\label{eq:cvv}
\bv^\epsilon&=\bv+\ii\epsilon\,\bbm=(v+\ii\epsilon)\bbm.
\end{align}
Indeed, upon introducing the dynamic elastic tensor \citep{SAEN53,TEUT62a}, modified with \eqref{eq:cvv} into a complex-valued tensor as
\begin{align}
\label{eq:modifiedc}
\widetilde{c}_{ijkl}^\epsilon &=c_{ijkl}-\rho v^\epsilon_j v^\epsilon_k\delta_{ij},
\end{align}
and the notation $(a b)^\epsilon_{il}=a_j\widetilde{c}_{ijkl}^\epsilon b_k$ \citep{ANDE17}, one has \citep{PELL17}
\begin{align}
\label{eq:gwithsign}
\sfG^+(\bk,\bk\cdot\bv)&=\lim_{\epsilon\to 0^+}\left[\Re+\ii\sign(\bhk\cdot\bbm)\Im\right]\frac{1}{k^2}{(\hk\hk)^\epsilon}^{-1},
\end{align}
the real and imaginary parts of which define the reactive and radiative parts of the Green operator, respectively (and, by inheritance, those of the fields that will be computed).\footnote{The concepts of reactive and radiative fields are borrowed from electrodynamics \citep[and references therein]{PELL17}. Elastic energy stored in a stationary manner at short distances from the source arises from the reactive field, whereas the radiative field (also called `active' field in acoustics) transports energy away from the source via wave propagation \citep{JACO89}. The underlying derivations are done in the time-harmonic regime with the acoustic Poynting vector $\bP=-\dot{\bu}\cdot\bsigma$ \citep{AULD73}, paralleling similar derivations in electrodynamics \citep{JACK98}.}

Expression \eqref{eq:gwithsign} allows for an evaluation of the angular integral over the direction $\bhk$ in \eqref{eq:betaB} by means of the Stroh formalism, while preserving the radiation features of the problem.

\subsection{Rescalings}
As far as regularization is concerned, the calculation can be made formally isotropic-like in the following manner. Introduce dimensionless position position and wave vectors $\br^z\equiv\sfz^{-1}\cdot\br\equiv(x^z,y^z)$, and $\bq=\sfz\cdot\bk$. Evidently, $x^z=x/a_\parallel$ and $y^z=y/a_\perp$. Letting $G^+_{jp}(\ldots)\equiv G^+_{jp}(\sfz^{-1}\cdot\bq,\bq\cdot\sfz^{-1}\cdot\bv)$ , this change of variables gives
\begin{align}
\label{eq:beta1}
\beta_{ij}(\br)
&=\frac{\ii\,b_l}{\det\sfz}z^{-1}_{ii'}\int\frac{\dd^2\!q}{(2\pi)^2}\left[q_{i'} G^+_{jp}(\ldots) q_{q'} z^{-1}_{q'q}c_{pqkl}-z_{i'k}\delta_{jl}\right]\frac{n_k \ee^{\ii \bq\cdot\br^z}\ee^{-q}}{\bq\cdot\bbm/a_\parallel+\ii 0^+}.
\end{align}
Upon introducing the $z$-rescaled elastic tensor
\begin{align}
\label{eq:zelast}
c^z_{ijkl}&=z^{-1}_{jj'}c_{ij'k'l}z^{-1}_{k'k},
\end{align}
and noting that $\sfz\cdot\bn=a_\perp \bn$ and $\det\sfz=a_\parallel a_\perp$, expression \eqref{eq:beta1} becomes
\begin{align}
\label{eq:beta2}
\beta_{ij}(\br)
&=\ii\,b_l z^{-1}_{ii'}\int\frac{\dd^2\!q}{(2\pi)^2}\left[q_{i'} G^+_{jp}(\ldots)(q_q c^z_{pqkl}n_k)-n_{i'}\delta_{jl}\right]\frac{\ee^{\ii \bq\cdot\br^z}\ee^{-q}}{\bq\cdot\bbm+\ii 0^+}.
\end{align}
Introduce now the notation $(ab)^{\epsilon z}$ ---the $z$-rescaled version of $(ab)^\epsilon$--- to denote the $3\times 3$ matrix of components
$(a b)^{\epsilon z}_{il}=a_jz^{-1}_{jj'}\widetilde{c}^\epsilon_{ij'k'l}z^{-1}_{k'k} b_k$.
From \eqref{eq:gwithsign} one deduces
\begin{align}
\label{eq:gwithsignz}
\sfG^+(\sfz^{-1}\cdot\bq,\bq\cdot\sfz^{-1}\cdot\bv)&=\lim_{\epsilon\to 0^+}\left[\Re+\ii\sign(\bhq\cdot\bbm)\Im\right]\frac{1}{q^2}{(\hq\hq)^{\epsilon z}}^{-1}.
\end{align}
The term within square brackets in \eqref{eq:beta2} does not depend on the modulus $q$, so that the integral over $q$ is done right away using
\begin{align}
\label{eq:intkiso}
\int_0^\infty \frac{\dd q\,q\, \ee^{\ii \bq\cdot\br^z}\ee^{-q}}{\bq\cdot\bbm+\ii 0^+}
=\frac{\ii}{(\bhq\cdot\bbm+\ii 0^+)(\bhq\cdot\br^z+\ii)}.
\end{align}
The remaining angular integral over $\bhq$ in \eqref{eq:beta2} is dealt with by introducing the angle $\phi$ such that $\bhq(\phi)=-\sin\phi\,\bbm+\cos\phi\,\bn$. By periodicity, the integral is equivalent to one over $\phi$. Introduce the shorthand notation $\av{f}_\phi=(2\pi)^{-1}\int_0^{2\pi}\dd\phi\,f(\phi)$ for angular averages. Substituting \eqref{eq:gwithsignz} and \eqref{eq:intkiso} into \eqref{eq:beta2}, and since $(\hat{q}_q c^z_{pqkl}n_k)$ is $(\hq n)^{\epsilon z}_{pl}$ because $\bv\cdot\bn=0$, one obtains
\begin{align}
\label{eq:beta3}
\beta_{ij}(\br)
&=-\frac{b_l}{2\pi}z^{-1}_{ii'}\lim_{\epsilon\to 0^+}\avphi{
\frac{\hq_{i'}\left[\Re+\ii\sign(\bhq\cdot\bbm)\Im\right][{(\hq\hq)^{\epsilon z}}^{-1}\cdot(\hq n)^{\epsilon z}]_{jl}-n_{i'}\delta_{jl}}{(\bhq\cdot\bbm+\ii 0^+)(\bhq\cdot\br^z+\ii)}
}.
\end{align}
Except for the leftmost application of $\sfz^{-1}$, the core anisotropy has been absorbed into the elastic constants and the redefinition of the position vector. The calculation becomes formally equivalent to one with an isotropic core of unit half width.

Hereafter, the limit $\epsilon\to 0^+$ is omitted and \emph{implicitly applies to all field-related expressions.} To implement this limit, numerical calculations such as in Sec.\ \ref{sec:nures} below, will be done using the finite tiny numerical value $\epsilon\simeq 10^{-10}$.

\subsection{Rotated basis, Stroh formalism, and angular integrals}
The next transformations closely follow Sec.\ 3.2  of \citet{PELL17} devoted to Model I (equations (71) to (77), and (82) to (85) in the latter reference, to which the reader is referred for details).

The unit vector $\bl(\phi)=\cos\phi\,\bbm+\sin\phi\,\bn$ is introduced to make $\{\bl,\bhq\}$ another orthonormal basis of the plane, which is the fixed basis $\{\bbm,\bn\}$ rotated by the angle $\phi$, and the unit vector $\bn$ is decomposed on this rotated basis. Focusing on the integrand in \eqref{eq:beta3}, and since $\bn\cdot\bl=-\bhq\cdot\bbm$, one finds
\begin{align}
&\frac{\hq_{i'}\left[\Re+\ii\sign(\bhq\cdot\bbm)\Im\right][{(\hq\hq)^{\epsilon z}}^{-1}\cdot(\hq n)^{\epsilon z}]_{jl}-n_{i'}\delta_{jl}}{\bhq\cdot\bbm+\ii 0^+}\nonumber\\
\label{eq:simplified}
&{}=-\left\{\hq_{i'}\Re[{(\hq\hq)^{\epsilon z}}^{-1}\cdot(\hq l)^{\epsilon z}]_{jl}-l_{i'}\delta_{jl}\right\}
-\ii\sign(\bhq\cdot\bbm)\hq_{i'}\Im[{(\hq\hq)^{\epsilon z}}^{-1}\cdot(\hq l)^{\epsilon z}]_{jl},
\end{align}
where a factor $(\bhq\cdot\bbm)$ has been eliminated out between the numerator and denominator. Next, the integrand in \eqref{eq:beta3} is the product of \eqref{eq:simplified} by
\begin{align}
\frac{1}{\bhq\cdot\br^z+\ii}=\frac{\bhq\cdot\br^z}{(\bhq\cdot\br^z)^2+1}-\frac{\ii}{(\bhq\cdot\br^z)^2+1}.
\end{align}
Separating the real and imaginary parts of this product, one observes that the imaginary terms are odd under the inversion symmetry $\phi\to\phi+\pi$, i.e., $\bhq\to-\bhq$ and $\bl\to-\bl$, and so do not contribute to the angular integral in \eqref{eq:beta3}. Thus, the distortion becomes
\begin{align}
\beta_{ij}(\br)
&=\frac{b_l}{2\pi}z^{-1}_{ii'}\Biggl[\Re\avphi{
\frac{\bigl\{\hq_{i'}[{(\hq\hq)^{\epsilon z}}^{-1}\cdot(\hq l)^{\epsilon z}]_{jl}-l_{i'}\delta_{jl}\bigr\}(\bhq\cdot\br^z)}{(\bhq\cdot\br^z)^2+1}
}\nonumber\\
\label{eq:beta4}
&\hspace{3cm}{}+\Im\avphi{
\frac{\sign(\bhq\cdot\bbm)\hq_{i'}[{(\hq\hq)^{\epsilon z}}^{-1}\cdot(\hq l)^{\epsilon z}]_{jl}}{(\bhq\cdot\br^z)^2+1}
}\Biggr].
\end{align}
At this point, we appeal to the Stroh identity
\begin{align}
\label{eq:stroh}
{(\hq\hq)^{\epsilon z}}^{-1}\cdot(\hq l)^{\epsilon z}&=-\sum_{\alpha=1}^6 p^\alpha(\phi)\bA^\alpha\otimes \bL^\alpha,
\end{align}
where the 3-vectors $\bA^\alpha$ and $\bL^\alpha$ are read, respectively, from the first three and last three components of the complex-valued 6-vectors $\zeta^\alpha=(A^\alpha_1,A^\alpha_2,A^\alpha_3,L^\alpha_1,L^\alpha_2,L^\alpha_3)$, $\alpha=1,\ldots,6$, which are the eigenvectors of the $6\times 6$ modified Stroh matrix (non-Hermitian)
\begin{align}
\label{eq:strohmat}
\mathcal{N}&=-
\left(
\begin{array}{cc}
(nn)_\epsilon^{z-1}\cdot(nm)^{\epsilon z}&{(nn)^{\epsilon z}}^{-1}\\
(mn)^{\epsilon z}\cdot{(nn)^{\epsilon z}}^{-1}\cdot(nm)^{\epsilon z}-(mm)^{\epsilon z}&\qquad
(mn)^{\epsilon z}\cdot{(nn)^{\epsilon z}}^{-1}
\end{array}
\right).
\end{align}
The eigenvectors are normalized such that $\bA^\alpha\cdot\bL^\beta+\bA^\beta\cdot\bL^\alpha=\delta_{\alpha\beta}$ \citep{ANDE17}. The key point of the Stroh formalism is that those vectors do not depend on the angle $\phi$. The dependence on $\phi$ in the right-hand side of \eqref{eq:stroh} solely arises from the functions $p^\alpha(\phi)\equiv \tan(\psi_\alpha-\phi)$, where the $\psi_\alpha$ are complex-valued angles. The eigenvalues of $\mathcal{N}$ are the quantities $p_\alpha\equiv p^\alpha(\phi=0)=\tan(\psi_\alpha)$ \citep{CHAD77}. With respect to the classical Stroh formalism modifications in the present implementation are twofold: a) the Stroh matrix is no more real-valued, as it depends on the infinitesimal imaginary number $\ii\epsilon$, which encodes the radiation condition \citep{PELL17}; (b) the modified elastic tensor in the $3\times 3$ matrices $(ab)^{\epsilon z}$ has been rescaled according to \eqref{eq:zelast}. For conciseness, the dependence of $p^\alpha$, $\bA^\alpha$, and $\bL^\alpha$ on $\epsilon$ and on $\sfz$ is left implicit. We emphasize that, quite generally, the modified Stroh formalism preserves \emph{all} the tensor identities (such as \eqref{eq:stroh}) of the classical one \citep{ANDE17}.

Substituting \eqref{eq:stroh} into \eqref{eq:beta4} yields an expression for the elastic distortion in terms of manageable angular integrals. Separating it into its reactive and radiative parts,
\begin{subequations}
\label{eq:betaexpr}
\begin{align}
\label{eq:betaexpr1}
\beta_{ij}(\br)&=\beta^{\rm react}_{ij}(\br)+\beta^{\rm rad}_{ij}(\br),\\
\label{eq:betaexpr2}
\beta^{\rm react}_{ij}(\br)&=-\frac{1}{2\pi}z^{-1}_{ii'}\left[
\Re\sum_\alpha
\avphi{
\frac{(\bhq\cdot\br^z)p^\alpha(\phi)\hq_{i'}}{(\bhq\cdot\br^z)^2+1}}
 A^\alpha_j(\bL^\alpha\cdot\bb)
+\avphi{\frac{(\bhq\cdot\br^z)l_{i'}}{(\bhq\cdot\br^z)^2+1}}b_j\right],\\
\label{eq:betaexpr3}
\beta^{\rm rad}_{ij}(\br)&=
-\frac{1}{2\pi}z^{-1}_{ii'}\Im\sum_\alpha\avphi{
\frac{\sign(\bhq\cdot\bbm)p^\alpha(\phi)\hq_{i'}}{(\bhq\cdot\br^z)^2+1}
}A^\alpha_j(\bL^\alpha\cdot\bb).
\end{align}
\end{subequations}

The integrals are computed as follows. Introduce the angle $\gamma$ such that $\br^z=r^z(\cos\gamma\bbm+\sin\gamma\bn)$. Then, $\bhq\cdot\bhr^z=\sin(\gamma-\phi)$. Introduce moreover $\varepsilon=1/r^z$, and the unit vector $\bhs^z=(-\sin\gamma,\cos\gamma)$ orthogonal to $\bhr^z$. Dropping for brevity the label $\alpha$ in $p^\alpha(\phi)$ and $\psi_\alpha$, and on account of the expression $p(\phi)=\tan(\psi-\phi)$ one has, with components expressed in the $\{\bbm,\bn\}$ basis,
\begin{subequations}
\label{eq:integrals}
\begin{align}
\label{eq:integral1}
\avphi{\frac{(\bhq\cdot\br^z)\bl}{(\bhq\cdot\br^z)^2+1}}
&=\varepsilon\avphi{\frac{\sin(\gamma-\phi)}{\sin^2(\gamma-\phi)+\varepsilon^2}
\left(
\begin{array}{c}
\cos\phi\\
\sin\phi
\end{array}
\right)},\\
\label{eq:integral2}
\avphi{
\frac{(\bhq\cdot\br^z)p(\phi)\bhq}{(\bhq\cdot\br^z)^2+1}}
&=
\varepsilon\avphi{\frac{\sin(\gamma-\phi)\tan(\psi-\phi)}{\sin^2(\gamma-\phi)+\varepsilon^2}
\left(
\begin{array}{c}
-\sin\phi\\
\cos\phi
\end{array}
\right)},\\
\label{eq:integral3}
\avphi{\frac{\sign(\bhq\cdot\bbm)p(\phi)\bhq}{(\bhq\cdot\br^z)^2+1}}
&=
\varepsilon^2\avphi{\frac{\sign(\sin\phi)\tan(\psi-\phi)}{\sin^2(\gamma-\phi)+\varepsilon^2}
\left(
\begin{array}{c}
\sin\phi\\
-\cos\phi
\end{array}
\right)}.
\end{align}
\end{subequations}

These integrals are markedly different from the ones for model I in \citet{PELL17}.\footnote{The counterparts in the latter reference of the present Eqs.\ \eqref{eq:integral1}, \eqref{eq:integral2} and \eqref{eq:integral3} are Eqs.\ (80b), (80a), and (86), respectively.} The main complication here resides in the factor $\sign(\sin\phi)$ in \eqref{eq:integral3}, which requires special care. The integrals are done in the Appendix by contour integration in the complex plane. The result is
\begin{subequations}
\label{eq:integralsresult}
\begin{align}
\label{eq:integral1result}
\avphi{\frac{(\bhq\cdot\br^z)\bl}{(\bhq\cdot\br^z)^2+1}}
&=-\varepsilon\left(1-\frac{\varepsilon}{\sqrt{1+\varepsilon^2}}\right)\bhs^z,\\
\avphi{
\frac{(\bhq\cdot\br^z)p(\phi)\bhq}{(\bhq\cdot\br^z)^2+1}}
&=\varepsilon
\left(1-\frac{\varepsilon}{\sqrt{1+\varepsilon^2}}\right)\bhs^z\nonumber\\
\label{eq:integral2result}
&\hspace{-2cm}{}+\varepsilon\left[
\frac{\varepsilon}{\sqrt{1+\varepsilon^2}}(\sin\gamma-p\cos\gamma)
+\ii s_\alpha(\cos\gamma+p\sin\gamma)
\right]\frac{\bbm+p\,\bn}{D_{\gamma\varepsilon}(p)},
\end{align}
\begin{align}
\avphi{\frac{\sign(\bhq\cdot\bbm)p(\phi)\bhq}{(\bhq\cdot\br^z)^2+1}}
&=
\varepsilon
\left[F^{(2)}(\br^z)\,\bhr^z+F^{(3)}(\br^z)\,\bhs^z\right]\nonumber\\
\label{eq:integral3result}
&\hspace{-3.7cm}{}+\varepsilon
\left[
\varepsilon F^{(1)}(p)-F^{(2)}(\br^z)(\cos\gamma+p\,\sin\gamma)
-F^{(3)}(\br^z)(\sin\gamma-p\,\cos\gamma)
\right]\frac{\bbm+p\,\bn}{D_{\gamma\varepsilon}(p)},
\end{align}
\end{subequations}
where $s_\alpha=\sign\Im p_\alpha$, and the following quantities have been introduced:
\begin{subequations}
\begin{align}
\label{eq:ddef}
D_{\gamma\varepsilon}(p)&=(\cos\gamma+p\sin\gamma)^2+(1+p^2)\varepsilon^2,\\
\label{eq:subeq1}
F^{(1)}(p)&=\sqrt{1+p^2}\left[\ii\,\sign(\Im p)-\frac{2}{\pi}{\rm Arcsinh}(p)\right],\\
\label{eq:subeq2}
F^{(2)}(\br^z)
&=\frac{1}{\pi}{\rm Arg}\left((\varepsilon+\ii\sin\gamma)^2\right)
=\frac{1}{\pi}{\rm Arg}\left((1+\ii y^z)^2\right),\\
\label{eq:subeq3}
F^{(3)}(\br^z)
&=\frac{2\,\varepsilon}{\pi\sqrt{1+\varepsilon^2}}{\rm Arctanh}\left(\frac{\cos\gamma}{\sqrt{1+\varepsilon^2}}\right)
=\frac{2}{\pi\sqrt{1+{r^z}^2}}{\rm Arctanh}\left(\frac{x^z}{\sqrt{1+{r^z}^2}}\right).
\end{align}
\end{subequations}

\subsection{The elastic distortion: final result}
\label{sec:edfr}
Upon substituting Eqs.\ \eqref{eq:integralsresult} into Eqs.\ \eqref{eq:betaexpr}, two notable simplifications arise: 1) the last term within brackets in \eqref{eq:betaexpr2} is eliminated, because of the completeness identity $\sum_\alpha\bA^\alpha\otimes\bL^\alpha=\sfI$ and of mutual cancellation between \eqref{eq:integral1result} and the first term of \eqref{eq:integral2result}; 2) the first term of \eqref{eq:integral3result} does not contribute under the $\Im$ operator of \eqref{eq:betaexpr3} because it is real-valued.

Replacing $\varepsilon$ by its expression $1/r^z$, the final result for the elastic distortion in Model III thus reads
\begin{subequations}
\label{eq:betaexprfin}
\begin{align}
\label{eq:betaexpr1fin}
&\beta_{ij}(\br)=\beta^{\rm react}_{ij}(\br)+\beta^{\rm rad}_{ij}(\br),\\
\label{eq:betaexpr2fin}
&\beta^{\rm react}_{ij}(\br)
=-\frac{1}{2\pi}z^{-1}_{ii'}
\Re\sum_\alpha
\left[
\ii s_\alpha(x^z+p_\alpha\,y^z)
+\frac{y^z-p_\alpha\,x^z}{\sqrt{r^{z\,2}+1}}
\right]
\frac{m_{i'}+p_\alpha\,{n_{i'}}}{\Delta_\alpha(\br^z)}
A^\alpha_j(\bL^\alpha\cdot\bb),\\
\label{eq:betaexpr3fin}
&\beta^{\rm rad}_{ij}(\br)
=
-\frac{1}{2\pi}z^{-1}_{ii'}\Im\sum_\alpha
\Bigl[
F^{(1)}(p_\alpha)-F^{(2)}(\br^z)(x^z+p_\alpha\,y^z)
\nonumber\\
&\hspace{3.8cm}{}
-F^{(3)}(\br^z)(y^z-p_\alpha\,x^z)
\Bigr]
\frac{m_{i'}+p_\alpha\,{n_{i'}}}{\Delta_\alpha(\br^z)}
A^\alpha_j(\bL^\alpha\cdot\bb),
\end{align}
\end{subequations}
where $\Delta_\alpha(\br^z)\equiv\varepsilon^{-2}D_{\gamma\varepsilon}(p_\alpha)=(x^z+p_\alpha y^z)^2+(1+p_\alpha^2)$.

Again, the eigenvalues $p_\alpha$ and vectors $\bA^\alpha$ and $\bL^\alpha$ are computed here and below from the modified Stroh matrix \eqref{eq:strohmat}. With regard to the above results, one key point of the modified Stroh formalism is that even though some (or even all of the) $p_\alpha$s are real-valued for faster-than-wave velocities in the limit $\epsilon\to 0$ \citep{STRO62a}, the signs $s_\alpha=\sign\Im p_\alpha$, which are discontinuous functions, are determined by their values for $\epsilon$ infinitesimal but nonzero, and thus never vanish.

Let the eigenvalues $p_\alpha$ and the vectors $\bA^\alpha$ and $\bL^\alpha$ be labelled by the name of the model. In model I,
the eigenvalues are determined by the equation \citep{STRO62a,ANDE17,PELL17}
\begin{align}
\label{eq:det1}
\det\left\{(mm)^\epsilon+[(mn)^\epsilon+(nm)^\epsilon]p_{\rm I}+(nn)^\epsilon p_{\rm I}^2\right\}=0.
\end{align}
In Model III, the $z$-rescaling of the elastic tensor changes this equation into
\begin{align}
\det\left\{(mm)^\epsilon+\left(\frac{a_\parallel} {a_\perp}p_{\rm III}\right)[(mn)^\epsilon+(nm)^\epsilon]+\left(\frac{a_\parallel}{a_\perp}p_{\rm III}\right)^2(nn)^\epsilon\right\}=0,
\end{align}
where $(mm)^\epsilon$, $(nm)^\epsilon$ and $(mm)^\epsilon$ are the same matrices as in \eqref{eq:det1}. Thus, $p_{\rm III}=(a_\perp/a_\parallel) p_{\rm I}$. Moreover, one easily shows from the modified Stroh matrix \eqref{eq:strohmat} that $\bA^{\alpha}_{\rm III}=\sqrt{a_\parallel a_\perp}\bA^{\alpha}_{\rm I}$, and that $\bL^{\alpha}_{\rm III}=\bL^{\alpha}_{\rm I}/\sqrt{a_\parallel a_\perp}$. With these relationships, Eqs.\ \eqref{eq:betaexprfin} could as well be expressed in terms of the eigenvalues and eigenvectors of the `causal' Stroh formalism used to solve Model I. This is left to the reader for brevity.

\subsection{Self-stress and resolved self-stress}
By the generalized Hooke law, the self-stress of the dislocation is $\sigma_{ij}(\br)=c_{ijkl}\beta_{kl}(\br)$. As in the standard theory of a subsonic Volterra dislocation no further simplification of this quantity is possible. The situation changes for the resolved self-stress $\sigma(\br)=\smash{\bn\cdot\bsigma(\br)\cdot\bhb}$, which identifies with the self-stress part of the Peach-Koehler force divided by $b$ if $\bb$ remains confined to the slip plane \citep{BACO79}. To compute it \citep{ANDE17}, observe that
\begin{align}
\label{eq:intermediate}
&n_j c_{ijk'l}z_{k'k}^{-1}(m_k+p_\alpha n_k)A^\alpha_l=
n_j \widetilde{c}^\epsilon_{ijk'l}z_{k'k}^{-1}(m_k+p_\alpha n_k)A^\alpha_l\\
&=a_\perp n_j z_{jj'}^{-1} \widetilde{c}^\epsilon_{ij'k'l}z_{k'k}^{-1}(m_k+p_\alpha n_k)A^\alpha_l
=a_\perp [(nm)^{\epsilon z}+p_\alpha(nn)^{\epsilon z}]_{il}A^\alpha_l=-a_\perp L^\alpha_i,\nonumber
\end{align}
where the identity $\bL^\alpha=-[(nn)^{\epsilon z}+p_\alpha(nm)^{\epsilon z}]\cdot\bA^\alpha$, which holds as well in the modified theory, has been used in the last step. From Eqs.\ \eqref{eq:betaexprfin} and \eqref{eq:intermediate}, we deduce
\begin{subequations}
\label{eq:resolvedstress}
\begin{align}
\label{eq:sigresolved}
\sigma(\br)&=\sigma^{\text{react}}(\br)+\sigma^{\text{rad}}(\br),\\
\label{eq:sigreact}
\sigma^{\text{react}}(\br)
&=\frac{a_\perp b}{2\pi}\Re\sum_{\alpha=1}^6 \left[\ii s_\alpha(x^z+p_\alpha y^z)+\frac{y^z-p_\alpha x^z}{\sqrt{{r^z}^2+1}}\right]\frac{(\bL^\alpha\cdot\bhb)^2}{\Delta_\alpha(\br^z)},\\
\sigma^{\text{rad}}(\br)&=
\frac{a_\perp b}{2\pi}\Im\sum_\alpha
\Bigl[
F^{(1)}(p_\alpha)-(x^z+p_\alpha\,y^z)F^{(2)}(\br^z)\nonumber\\
\label{eq:sigrad}
&\hspace{3.6cm}{}-(y^z-p_\alpha\,x^z)F^{(3)}(\br^z)\Bigr]\frac{(\bL^\alpha\cdot\bhb)^2}{\Delta_\alpha(\br^z)}.
\end{align}
\end{subequations}

\subsection{Radiative resistance to motion}
The dissipative (self-)force experienced by the dislocation at fixed core width is obtained by projecting the resolved stress onto the dislocation density \citep{ESHE53b,PELL12,PELL14}. Using Parseval's theorem, this is
\begin{align}
\label{eq:fdiss}
f^{\rm diss}&=\int\dd^2r\,\rho(\br)\sigma(\br)=\int\frac{\dd^2k}{(2\pi)^2}\,\rho(-\bk)\sigma(\bk)=b\,\sigma(\br=\mathbf{0})|_{\sfz\to2\sf z},
\end{align}
where $\sigma(\br)$ is given by \eqref{eq:sigresolved}. The prescription in the rightmost expression indicates that the $\sfz$ matrix must be replaced by twice its value for the elastic fields. This last result is obvious from delineating the relationship between $\sigma(\br)$ and $\beta_{ij}(\br)$, going back to definition \eqref{eq:betaB} of the latter in terms of Fourier transforms, and using
\begin{align}
\rho(-\bk)\ee^{-\sqrt{\bk\cdot\sfz^2\cdot\bk}}&=b\,\ee^{-2\sqrt{\bk\cdot\sfz^2\cdot\bk}}=b\,\ee^{-\sqrt{\bk\cdot(2\sfz)^2\cdot\bk}}.
\end{align}
This convenient property stems from the regularization method employed. Equation \eqref{eq:sigreact} shows that the reactive component of the self-stress vanishes at $\br=\mathbf{0}$, so that the self-force is purely radiative. Substituting \eqref{eq:sigrad} into \eqref{eq:fdiss} yields
\begin{align}
\label{eq:fdissexplicit}
f^{\rm diss}
&=b\,\sigma^{\rm rad}(\br=\mathbf{0})|_{\sfz\to2\sf z}=\frac{a_\perp}{2\pi}\Im\sum_\alpha \frac{F^{(1)}(p_\alpha)}{1+p_\alpha^2}\left.(\bL^\alpha\cdot\bb)^2\right|_{\sfz\to2\sf z}.
\end{align}

\section{The isotropic-core limit of model II}
\label{sec:isolim}
For an isotropic core with $a_\parallel=a_\perp=a$ (i.e., $\sfz=a\,\sfI$) immediate simplifications of \eqref{eq:betaexpr2fin} and \eqref{eq:betaexpr3fin} yield, with now $\Delta_\alpha(\br)\equiv (x+p_\alpha\,y)^2+(1+p_\alpha^2)a^2$, the following expression where, due to the relationships discussed in Sec.\ \ref{sec:edfr}, the Stroh eigenvalues and eigenvectors of Model I can be used indifferently:
\begin{subequations}
\label{eq:betaexpriso}
\begin{align}
\label{eq:betaexpr2iso}
\beta^{\rm react}_{ij}(\br)
&=-\frac{1}{2\pi}\Re\sum_\alpha
\left[
\ii s_\alpha(x+p_\alpha\,y)
+\frac{a(y-p_\alpha\,x)}{\sqrt{r^2+a^2}}
\right]
\frac{m_{i}+p_\alpha\,{n_{i}}}{\Delta_\alpha(\br)}
A^{\alpha}_j(\bL^{\alpha}\cdot\bb),\\
\beta^{\rm rad}_{ij}(\br)
&=
-\frac{1}{2\pi}\Im\sum_\alpha
\Bigl[
a F^{(1)}(p)-(x+p_\alpha\,y)F^{(2)}\!\!\left(\frac{\br}{a}\right)
\nonumber\\
\label{eq:betaexpr3iso}
&\hspace{3cm}{}
-(y-p_\alpha\,x)F^{(3)}\!\!\left(\frac{\br}{a}\right)
\Bigr]
\frac{m_{i}+p_\alpha\,{n_{i}}}{\Delta_\alpha(\br)}
A^{\alpha}_j(\bL^{\alpha}\cdot\bb).
\end{align}
\end{subequations}

\section{Numerical illustrations}
\label{sec:nures}
The theory is illustrated for an edge dislocation in b.c.c.\ Fe, on the slip system $\bbm=(1,1,1)/\sqrt{3}$, $\bn=(1,-1,0)/\sqrt{2}$. The elastic constants (in GPa) are $C_{11}=226$, $C_{12}=140$, and $C_{44}=116$, the material density is $\rho=7.8672$ g/cm${}^3$, and the lattice parameter is $a_0=0.287$ nm \citep{LIDE10}. All lengths are measured in units of the interplanar distance $d=a_0/\sqrt{2}$, and the stress is measured in units of the theoretical shear stress $\sigma_{\rm th}=\mu\,b/(2\pi d)$, where $\mu=m_i n_j c_{ijkl} n_k m_l=13.125$ GPa is the in-plane shear modulus. The Burgers vector is $\bb=(\sqrt{3}/2)a_0\bbm$. In this section, the transverse half-width is always $a_\perp=d/2$, which is a reasonable assumption.

\begin{figure}
\centerline{\includegraphics[width=12.5cm]{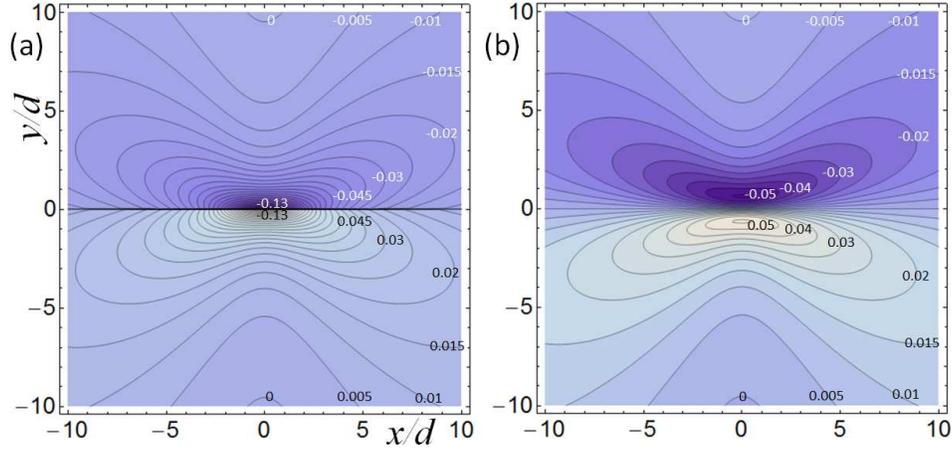}}
\vspace*{3pt}
\caption{\label{fig:fig1} Comparison between Model I and Model III for $a_\parallel= 5 a_\perp$. Dimensionless stress component $\sigma_{12}/\sigma_{\rm th}$ at $v=2000$ m/s. a) Model I \citep{PELL17} with $a=a_\parallel$; b) Model III.}
\end{figure}
Fig.\ \ref{fig:fig1} shows in the subsonic regime the effect of a finite $a_\perp$ in Model III, compared with the Peierls-Eshelby dislocation of Model I where $a_\perp=0$ by construction: fields become continuous, and are lower near the core center in Model III.

\begin{figure}
\centerline{\includegraphics[width=12.5cm]{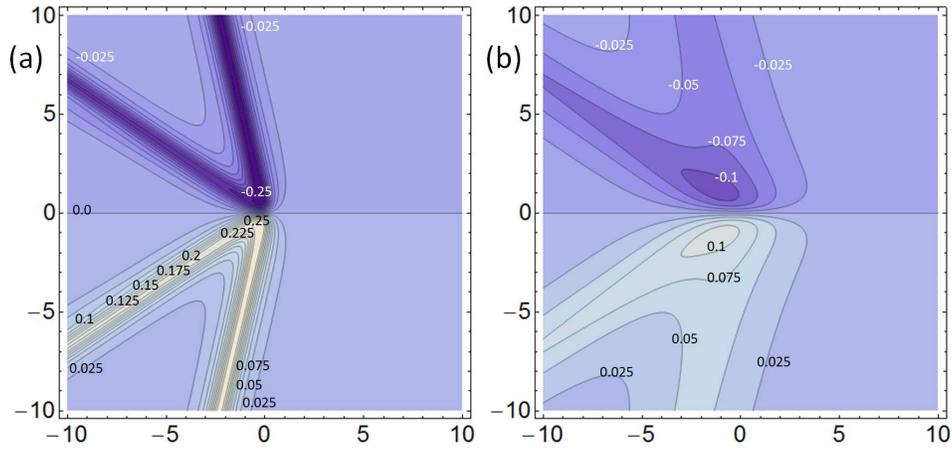}}
\vspace*{3pt}
\caption{\label{fig:fig2} Effect of a larger core width in Model III in the intersonic range. Dimensionless stress component $\sigma_{12}/\sigma_{\rm th}$ at $v=3500$ m/s. a) $a_\parallel=a_\perp$ (isotropic-core limit, i.e., Model II). b) $a_\parallel= 5 a_\perp$.}
\end{figure}

Fig.\ \ref{fig:fig2} illustrates the influence of the parallel core width $a_\parallel$ in Model III, in the intersonic range, at fixed $a_\perp$. The dislocation velocity is larger than two of the wavespeeds for the geometry considered, so that two Mach cones are present. The local intensity of the fields is drastically reduced when the core width is larger, especially within the Mach cones, which become wider.

The angle of the the Mach cone branches is unchanged, as must be. The reason is that the geometric equation for each cone branch is $x+p_\alpha y=0$, where the  eigenvalue $p_\alpha$ relative to the cone branch considered is real-valued in the supersonic range in the limit $\epsilon\to 0$. The equation for the cone branches in Model III is $x^z+p_{\rm III}\,y^z=0$. Since $p_{\rm III}=(a_\perp/a_\parallel)p_{\rm I}$, and $x^z=x/a_\parallel$ and $y^z=y/a_\perp$, this equation is the same as for Model I.

\begin{figure}
\centerline{\includegraphics[width=10cm]{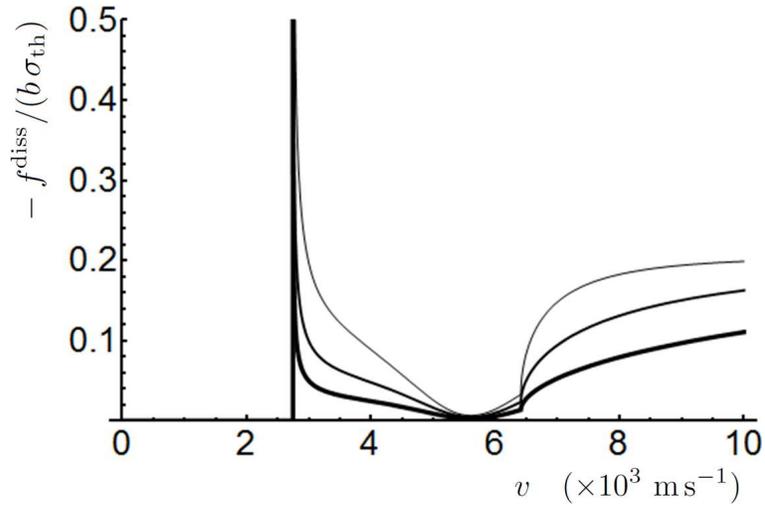}}
\vspace*{3pt}
\caption{\label{fig:fig3} Normalized radiative dissipation force vs.\ velocity $v$ for different velocity-independent core widths $a_\parallel=a_\perp$, $2a_\perp$, and $4a_\perp$. The thickness of the plot increases with $a_\parallel$.}
\end{figure}
Fig.\ \ref{fig:fig3} displays the dependence on the fixed core width $a_\parallel$ of the dissipative force $f^{\rm diss}$, Equ.\ \eqref{eq:fdissexplicit}, which is of sign opposite to $v$. The overall dissipation decreases as the longitudinal core width increases. The figure illustrates several other aspects of the problem, among which the critical velocities $v_1=2.75\times 10^3$ m s$^{-1}$ at which the intersonic regime (with at least one Mach cone) begins, and $v_3=6.41 \times 10^3$ m s$^{-1}$, at which the fully supersonic regime (with three Mach cones) begins; see \citet{PELL17} for wave patterns. Notable variations of the dissipation occurs at these special velocities. The dissipation is zero in the subsonic range $v<v_1$, where the self-force does no work \citep{ANDE17}. Indeed, a detailed study of Eqs.\ \eqref{eq:betaexprfin} would show that $\beta_{ij}(\br=\mathbf{0})$ vanishes in this regime. By contrast, even though $\sigma_{12}(\br=\mathbf{0})$ vanishes for all velocities (e.g., Figs.\ \ref{fig:fig1} and \ref{fig:fig2}), other components of $\sigma_{ij}(\br=\mathbf{0})$ remain nonzero for $v>v_1$, which leads to finite dissipation via radiation in Mach cones. The force is minimal, of order $10^{-3} b\sigma_{\rm th}$ near to $v\simeq 5.6(1)\times 10^3$ m s${}^{-1}$. This minimizer is quasi-independent of the ratio $a_\parallel/a_\perp$. Thus, the configuration is such that a `radiation-free' velocity does not exist \citep{GAOH99,BARN02a}, but is replaced by a `least-radiation' velocity. We speculate that the concept of a `least-radiation' velocity is always applicable.

\section{Concluding discussion}
\label{sec:concl}
The proposed regularization allows for computations of the fields in an anisotropic medium, for a core with elliptical shape, in terms of closed-form expressions involving elementary functions and quantities accessible numerically from the Stroh formalism. The method delivers smooth, non-singular, field expressions that apply to any velocity, thus encompassing the supersonic domain at no additional price. The numerical results illustrate the dramatic effect of changing the core width on the intensity of the field, which is an important issue for dislocation-dynamics simulations, especially with regard to field-triggered nucleation processes. Obtaining the radiative force is immediate if the core width is fixed. This force is a fundamental piece of the steady-state stress-velocity law for dislocations \citep{ROSA01}. However, an ingredient is missing to complete the present theory, namely, the dependence in the velocity of the longitudinal core width $a_\parallel$. In principle, such a function $a_\parallel(v)$ could be deduced from a steady-state version of the method of \citet{PELL14} to derive equations of motion for dislocations. A current physical limitation of the approach is that the functional form of the dislocation density is postulated rather than computed (in particular, this form might not be fully suited to supersonic motion). This should be contrasted with the markedly different field theory with gradient elasticity and quadratic core energy term \citep{LAZA09a}, which yields an equation for the dislocation density with Lorentz-Fitzgerald core contraction effects included. In this connection, the existence of Eq.\ \eqref{eq:diffeq} to produce the postulated dislocation density might provide a bridge between both approaches. These issues are left to future work.
\section*{Acknowledgments}
The author thanks his colleagues A.\ Vattr\'e and R.\ Madec for discussions.
\appendix
\noindent
\section{Angular integrals}
\label{sec:angint}
\subsection{Integrals \eqref{eq:integral1} and \eqref{eq:integral2}}
The integrands are continuous functions of $\phi$, and the integrals are computed componentwise by contour integration on $\Gamma$, the unit circle $|z|=1$, by means of the change of variables $z=\ee^{\ii\phi}$, and the formula
\begin{align}
\label{eq:cform1}
\int_0^{2\pi}\frac{\dd\phi}{2\pi}f(\phi)=\frac{1}{2\ii\pi}\int_{\Gamma}\frac{\dd z}{z}f(-\ii\log z)=\sum_{\text{poles }|z_k|<1}\text{Res }g(z)|_{z=z_k},
\end{align}
where $g(z)=f(-\ii\log z)/z$. The integrands in integrals \eqref{eq:integral1} have 5 poles $z_k$, $k=1,\ldots,5$, and those in \eqref{eq:integral2} have two more poles $z_{6,7}$. These are
\begin{align}
z_1&=0,
z_{2,3}=\pm\ee^{\ii\gamma}\left(\sqrt{1+\varepsilon^2}-\varepsilon\right),
z_{4,5}=\pm\ee^{\ii\gamma}\left(\sqrt{1+\varepsilon^2}+\varepsilon\right),
z_{6,7}=\pm\sqrt{\frac{p-\ii}{p+\ii}},
\end{align}
where $p=\tan\psi$. Thus, $|z_{2,3}|<1$, $|z_{4,5}|>1$, and $|z_{6,7}|<1$ if $\Im p >0$, and $|z_{6,7}|>1$ if $\Im p <0$.

For integral \eqref{eq:integral1} the integrand reads
\begin{align}
\label{eq:integrand1}
\frac{1}{z}\frac{\sin(\gamma-\phi)}{\sin^2(\gamma-\phi)+\varepsilon^2}
\left(
\begin{array}{c}
\cos\phi\\
\sin\phi
\end{array}
\right)
&=
-\ii
\frac{\ee^{\ii\gamma}\left(z^2-\ee^{2\ii\gamma}\right)}
{\prod_{k=1}^5(z-z_k)}
\left(
\begin{array}{c}
1+z^2\\
\ii(1-z^2)
\end{array}
\right)\equiv g(z).
\end{align}
Only the poles $z_{1,2,3}$ contribute to the result. Residues are denoted as $R^m_k$ for the $\bbm$ component and $R^n_k$ for the $\bn$ component. The shorthand notation $X=\varepsilon/\sqrt{1+\varepsilon^2}$ is used hereafter. From \eqref{eq:integrand1} one gets
\begin{subequations}
\begin{align}
R_1^m&=\ii(\cos\gamma-\ii\sin\gamma),\qquad R^m_{2,3}=-\frac{\ii}{2}\left(\cos\gamma-\ii\,X\sin\gamma\right),\\
R_1^n&=-(\cos\gamma-\ii\sin\gamma),\qquad R^n_{2,3}=\frac{1}{2}\left(X\cos\gamma-\ii\sin\gamma\right).
\end{align}
\end{subequations}
By \eqref{eq:cform1} the results are of the type $R_1+R_2+R_3=R_1+2R_2$ for both components, which leads to \eqref{eq:integral1result}.

For integral \eqref{eq:integral2} the integrand reads
\begin{align}
&\frac{1}{z}\frac{\sin(\gamma-\phi)\tan(\psi-\phi)}
{\sin^2(\gamma-\phi)+\varepsilon^2}
\left(
\begin{array}{c}
-\sin\phi\\
\cos\phi
\end{array}
\right)\nonumber\\
\label{eq:integrand2}
&\hspace{2cm}{}=
\frac{\ee^{\ii\gamma}\left(z^2-\ee^{2\ii\gamma}\right)}
{\prod_{k=1}^{7}(z-z_k)}
\left(z^2+\frac{p-\ii}{p+\ii}\right)
\left(
\begin{array}{c}
-\ii(1-z^2)\\
1+z^2
\end{array}
\right)\equiv g(z).
\end{align}
Only the poles $z_{1,2,3,6,7}$ contribute. The residues deduced from \eqref{eq:integrand2} are
\begin{subequations}
\begin{align}
R^m_1&=-\ii(\cos\gamma-\ii\sin\gamma),
\qquad
R^m_{6,7}=\frac{\ii(\cos\gamma+p\sin\gamma)}{D_{\gamma\varepsilon}(p)},\\
R^m_{2,3}&=
\frac{\ii}{2}\left(\cos\gamma-\ii\,X\sin\gamma\right)+\frac{-(p X+\ii)\cos\gamma+(X-\ii\,p)\sin\gamma}{2D_{\gamma\varepsilon}(p)},
\end{align}
and
\begin{align}
R^n_1&=\ii\,R^m_1,\qquad R^n_{6,7}=p\,R^m_{6,7} \\
R^n_{2,3}&=
-\frac{1}{2}\left(X\cos\gamma-\ii\,\sin\gamma\right)+p\frac{-(p X+\ii)\cos\gamma+(X-\ii\,p)\sin\gamma}{2D_{\gamma\varepsilon}(p)},
\end{align}
\end{subequations}
where $D_{\gamma\varepsilon}(p)$ is defined in \eqref{eq:ddef}. By \eqref{eq:cform1}, and given the locations of the poles, the results are of the type $R_1+R_2+R_3=R_1+2R_2$ if $\Im p<0$, and $R_1+R_2+R_3+R_6+R_7=R_1+2R_2+2R_6$ if $\Im p>0$, which can be encoded as $R_1+2R_2+R_6+\sign(\Im p)R_6$. Immediate simplifications then lead to \eqref{eq:integral2result}.

\subsection{Integral \eqref{eq:integral3}}
The integrand of \eqref{eq:integral3} is invariant under the symmetry $\phi\to\phi+\pi$.
The integral is folded onto the interval $[0,\pi]$ where $\sign(\sin\phi)=1$, which eliminates this factor. The change of variables $z=\tan(\phi/2)$  \citep{CART95} is used thereafter. Thus,
\begin{align}
&\avphi{\frac{\tan(\psi-\phi)\sign(\sin\phi)}{\sin^2(\gamma-\phi)+\varepsilon^2}
\left(
\begin{array}{c}
\sin\phi\\
-\cos\phi
\end{array}
\right)}
=\int_0^\pi\frac{\dd\phi}{\pi}\frac{\tan(\psi-\phi)}{\sin^2(\gamma-\phi)+\varepsilon^2}
\left(
\begin{array}{c}
\sin\phi\\
-\cos\phi
\end{array}
\right)\nonumber\\
\label{eq:transformedint}
&=\frac{1}{\pi}\int_0^{+\infty}\dd z
\frac{2[p(z^2-1)+2z]}
{(z^2-2p\,z-1)[(2z\cos\gamma+(z^2-1)\sin\gamma)^2+(1+z^2)^2\varepsilon^2]}
\left(
\begin{array}{c}
2z\\
z^2-1
\end{array}
\right).
\end{align}
\begin{figure}
\centerline{\includegraphics[width=4cm]{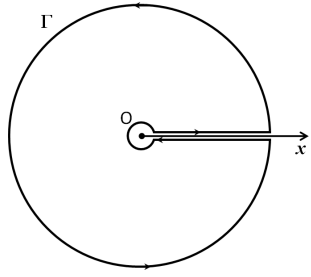}}
\vspace*{3pt}
\caption{\label{fig:fig4}Integration contour $\Gamma$ for integral \eqref{eq:transformedint}, where the contour radius goes to infinity.}
\end{figure}
Integral \eqref{eq:transformedint} can then be computed as \citep{CART95}
\begin{align}
\label{eq:icontour2}
\frac{1}{\pi}\int_0^{+\infty}\dd z\,f(z)&=-\frac{1}{\pi}\frac{1}{2\ii\pi}\int_{\Gamma}\dd z\,\log(-z)f(z)=-\frac{1}{\pi}\sum_{\text{poles }z_k}\mathop{\rm Res}f(z)_{z=z_k}\log(-z_k),
\end{align}
where $f(z)$ is the integrand in \eqref{eq:transformedint}, and where the new contour $\Gamma$ is depicted in Fig.\ \ref{fig:fig4}. The sum is over all the poles of $f(z)$ with nonzero imaginary part. The six poles of $f(z)$ are such, because in the modified Stroh formalism the $p_\alpha$ are always complex-valued due to the radiation condition; see Sec.\ \ref{sec:edfr}. They read
\begin{align}
z_{1,2}&=p\pm\sqrt{1+p^2},\quad
z_{3,4}=-\frac{\cos\gamma\pm\sqrt{1+\varepsilon^2}}{\sin\gamma+\ii\varepsilon},\quad z_{5,6}=-\frac{\cos\gamma\pm\sqrt{1+\varepsilon^2}}{\sin\gamma-\ii\varepsilon}.
\end{align}
Let $s_\gamma(p)=\sin\gamma-p\cos\gamma$ and $c_\gamma(p)=\cos\gamma+p\sin\gamma$. The corresponding residues for both components of $f(z)$ are
\begin{subequations}
\label{eq:i3mres}
\begin{align}
R^m_{1,2}&=\pm\frac{\sqrt{1+p^2}}{D_{\gamma\varepsilon}(p)},\qquad
R^n_{1,2}=\pm\frac{p\,\sqrt{1+p^2}}{D_{\gamma\varepsilon}(p)},
\end{align}
\begin{align}
R^m _{3,4,5,6}&=\pm\frac{1}{2\sqrt{1+\varepsilon^2}}
\left(\sin\gamma+\frac{s_\gamma(p)}{D_{\gamma\varepsilon}(p)}\right)
\pm\frac{\ii}{2\varepsilon}
\left(\cos\gamma-\frac{c_\gamma(p)}{D_{\gamma\varepsilon}(p)}\right),
\end{align}
\begin{align}
R^n_{3,4,5,6}&=\pm\frac{1}{2\sqrt{1+\varepsilon^2}}
\left(-\cos\gamma+p\frac{s_\gamma(p)}{D_{\gamma\varepsilon}(p)}\right)
\pm\frac{\ii}{2\varepsilon}
\left(\sin\gamma-p\frac{c_\gamma(p)}{D_{\gamma\varepsilon}(p)}\right),
\end{align}
\end{subequations}
where the residues $R_3$, $R_4$, $R_5$, and $R_6$ are defined, respectively, by the pairs of signs $(+,+)$, $(-,+)$, $(+,-)$, and $(-,-)$. Due to these alternating signs, the sum \eqref{eq:icontour2} involves the following combinations of logarithms, in which principal determinations of elementary functions are used:
\begin{subequations}
\begin{align}
&\log(-z_1)-\log(-z_2)=2{\rm Arcsinh}(p)-\ii\pi\sign(\Im p),\\
&\log(-z_3)-\log(-z_4)+\log(-z_5)-\log(-z_6)=4{\rm Arctanh}\left(\frac{\cos\gamma}{\sqrt{1+\varepsilon^2}}\right),\\
&\log(-z_3)+\log(-z_4)-\log(-z_5)-\log(-z_6)=2\ii{\rm Arg}\left((\varepsilon+\ii\sin\gamma)^2\right).
\end{align}
\end{subequations}
Re-organizing the sum \eqref{eq:icontour2} with the help of the above results and definitions \eqref{eq:subeq1}--\eqref{eq:subeq3} for each component immediately  gives expression \eqref{eq:integral3result}.


\markboth{Y.-P. Pellegrini}{Uniformly-moving non-singular dislocations with elliptical core shape in anisotropic media}
\newcommand{\enquote}[1]{``#1''}
\providecommand{\natexlab}[1]{#1}

\end{document}